\documentclass[%
reprint,
superscriptaddress,
 amsmath,amssymb,
 aps,
 8pt,
prb,
]{revtex4-2}
\usepackage{siunitx}

\usepackage{natbib}

\usepackage{graphicx}
\usepackage{dcolumn}
\usepackage{bm}
\usepackage{amssymb}
\usepackage{amsmath}
\usepackage{comment}

\usepackage{color}
\usepackage{hyperref}
\hypersetup{colorlinks=true, linkcolor=blue, citecolor=blue, urlcolor=blue,}

\newcommand{\C}[1]{\ensuremath{#1\,^\circ\text{C}}}

\usepackage{orcidlink}

\begin{document}

\setcounter{topnumber}{2}
\setcounter{dbltopnumber}{1}
\renewcommand{\topfraction}{0.85}
\renewcommand{\dbltopfraction}{0.9}
\renewcommand{\textfraction}{0.1}
\renewcommand{\floatpagefraction}{0.6}
\renewcommand{\dblfloatpagefraction}{0.6}
\setlength{\textfloatsep}{12pt plus 4pt minus 4pt}
\setlength{\dbltextfloatsep}{16pt plus 4pt minus 4pt}




\title{Defect Dynamics and Isomer Quenching in $^{229}$Th:CaF$_2$ Crystals}

\author{Ming Guan\orcidlink{0009-0000-3563-2516}}
\email{guanming@apm.ac.cn}
\affiliation{Research Institute for Interdisciplinary Science, Okayama University, Okayama, 700-8530, Japan}
\affiliation{Innovation Academy for Precision Measurement Science and Technology, Chinese Academy of Sciences, Wuhan 430071, China}


\author{Michael Bartokos\orcidlink{0009-0004-7610-9122}}
\affiliation{Institute for Atomic and Subatomic Physics, Atominstitut, TU Wien, 1020 Vienna, Austria}
\author{Kjeld Beeks\orcidlink{0000-0002-8707-6723}}
\affiliation{Institute for Atomic and Subatomic Physics, Atominstitut, TU Wien, 1020 Vienna, Austria}
\author{Takahiro Hiraki\orcidlink{0000-0002-6235-5830}}
\affiliation{Research Institute for Interdisciplinary Science, Okayama University, Okayama, 700-8530, Japan}
\author{Shinji Kitao\orcidlink{0000-0003-1457-9087}}
\affiliation{Institute for Integrated Radiation and Nuclear Science, Kyoto University, Osaka 590-0494, Japan}
\author{Takahiko Masuda\orcidlink{0000-0001-8122-5145}}
\affiliation{Research Institute for Interdisciplinary Science, Okayama University, Okayama, 700-8530, Japan}
\author{Nobumoto Nagasawa\orcidlink{0000-0001-5654-7708}}
\affiliation{Japan Synchrotron Radiation Research Institute,  Kouto,  Hyogo 679-5198, Japan}
\author{Koichi Okai}
\thanks{Present affiliation: Graduate School of Science, Osaka University, Toyonaka, Osaka 560-0043, Japan}
\affiliation{Research Institute for Interdisciplinary Science, Okayama University, Okayama, 700-8530, Japan}
\author{Ryoichiro Ogake}
\affiliation{Research Institute for Interdisciplinary Science, Okayama University, Okayama, 700-8530, Japan}
\author{Martin Pimon\orcidlink{https://orcid.org/0000-0001-7784-463X}}
\affiliation{Institute for Atomic and Subatomic Physics, Atominstitut, TU Wien, 1020 Vienna, Austria}

\author{Noboru Sasao\orcidlink{https://orcid.org/0000-0002-2685-7905}}
\affiliation{Research Institute for Interdisciplinary Science, Okayama University, Okayama, 700-8530, Japan}
\author{Fabian Schaden\orcidlink{0000-0001-7154-0440}}
\affiliation{Institute for Atomic and Subatomic Physics, Atominstitut, TU Wien, 1020 Vienna, Austria}
\author{Thorsten Schumm\orcidlink{0000-0002-1066-202X}}
\affiliation{Institute for Atomic and Subatomic Physics, Atominstitut, TU Wien, 1020 Vienna, Austria}
\author{Makoto Seto}
\affiliation{Institute for Integrated Radiation and Nuclear Science, Kyoto University, Osaka 590-0494, Japan}
\author{Kenji Tamasaku\orcidlink{0000-0003-1440-0518}}
\affiliation{RIKEN SPring-8 Center,  Kouto,  Hyogo 679-5148, Japan}
\author{Sayuri Takatori\orcidlink{0000-0002-8705-9624}}
\affiliation{Research Institute for Interdisciplinary Science, Okayama University, Okayama, 700-8530, Japan}
\author{Yoshitaka Yoda\orcidlink{https://orcid.org/0000-0003-3062-8651}}
\affiliation{Japan Synchrotron Radiation Research Institute,  Kouto,  Hyogo 679-5198, Japan}
\author{Akihiro Yoshimi\orcidlink{0000-0002-2438-1384}}
\affiliation{Research Institute for Interdisciplinary Science, Okayama University, Okayama, 700-8530, Japan}
\author{Koji Yoshimura\orcidlink{0000-0002-2415-718X}}
\affiliation{Research Institute for Interdisciplinary Science, Okayama University, Okayama, 700-8530, Japan}


\date{\today}

\begin{abstract}
Thorium-229 doped calcium fluoride ($^{229}$Th:CaF$_2$) is a leading candidate for a solid-state nuclear clock, owing to the unusually low energy of the $^{229}$Th isomer transition at 8.3557\,eV and the very high banggap of CaF$_2$ around 12\,eV. Here we report a temperature-dependent study of $^{229}$Th:CaF$_2$ luminescence, covering radioluminescence, photoluminescence, thermoluminescence, and afterglow. Using synchrotron x-ray excitation together with a cryo-vacuum system, we resolve the thermal quenching of the self-trapped exciton (STE) scintillation into its triplet and singlet components, identify eleven thermoluminescence glow peaks together with their associated emission bands, and separate the temperature-dependent components that make up the long-lived afterglow. The trap depths obtained in this way are closely correlated with the temperature dependence of x-ray-induced isomer quenching, including the enhanced isomer yield near \SI{-60}{\celsius} and the minimum near \SI{-80}{\celsius}, supporting the carrier-trapping picture in which trapped carriers are unavailable to quench the nucleus. Taken together, these results connect the optical and nuclear observables of $^{229}$Th:CaF$_2$ through a unified picture of electron defects and Th-related traps, while providing a comprehensive characterization of the material relevant to solid-state nuclear clock operation.
\end{abstract}






\maketitle
\section{Introduction}

The $^{229}$Th nucleus possesses a uniquely low-lying excited state at 8.3557\,eV, referred to as the thorium isomer $^{229\text{m}}$Th~\cite{kroger1976features, Beck2007, kraemer2023observation}. Because this transition can be driven by vacuum-ultraviolet (VUV) lasers, $^{229}$Th has become a prime candidate for a nuclear clock, which would in principle offer superior stability and a reduced sensitivity to environmental perturbations as compared with atomic clocks~\cite{peik2003nuclear}. Two routes towards such a clock are currently pursued: trapping $^{229}$Th$^{3+}$ ions \cite{peik2003nuclear, campbell2009multiply, campbell2012single, yamaguchi2024laser, scharl2023setup}, or doping $^{229}$Th$^{4+}$ into a VUV-transparent crystal \cite{rellergert2010progress, kazakov2012performance, gong2024feasibility}. In the latter approach, usually termed the solid-state nuclear clock, a large number of nuclei are exploited to suppress quantum projection noise, and laser cooling is not required~\cite{kazakov2012performance}. Among the wide-bandgap hosts that have been considered, Th:CaF$_2$ is by far the most extensively studied \cite{kazakov2012performance, nickerson2020nuclear, nickerson2021driven}.


Solid-state nuclear clocks based on thorium-229-doped calcium fluoride ($^{229}$Th:CaF$_2$) have recently been operated in three laboratories, with their transition frequencies compared across the three systems \cite{de2026thorium, huang2026nuclear, ooi2025frequency}. Of these, the two these generate the clock error signal by absorption spectroscopy~\cite{de2026thorium, huang2026nuclear}, while the third reads out the nuclear state through fluorescence~\cite{ooi2025frequency}. Alongside these advances, the interaction between the Th dopant and the host lattice remains comparatively poorly understood. Earlier studies treated the CaF$_2$ crystal essentially as a static holder that does not perturb the nucleus \cite{tiedau2024laser, hiraki2024controlling, zhang2024frequency}. This picture is incomplete, because electrons and holes in CaF$_2$ are mobile, and crystal defects can capture them as color centers that subsequently recombine and emit photons; CaF$_2$ therefore constitutes a dynamic environment for the nucleus. Consistent with this view, recent isomer quenching experiments reveal a non-negligible coupling between the nucleus and the surrounding matrix~\cite{schaden2025laser, terhune2025photo, guan2025x}, and controlling both the excitation and the decay therefore requires that this coupling be understood. In this work we report luminescence measurements of Th:CaF$_2$ that elucidate the underlying defect dynamics.

\begin{figure*}[!t]
    \centering
    \includegraphics[width=\linewidth]{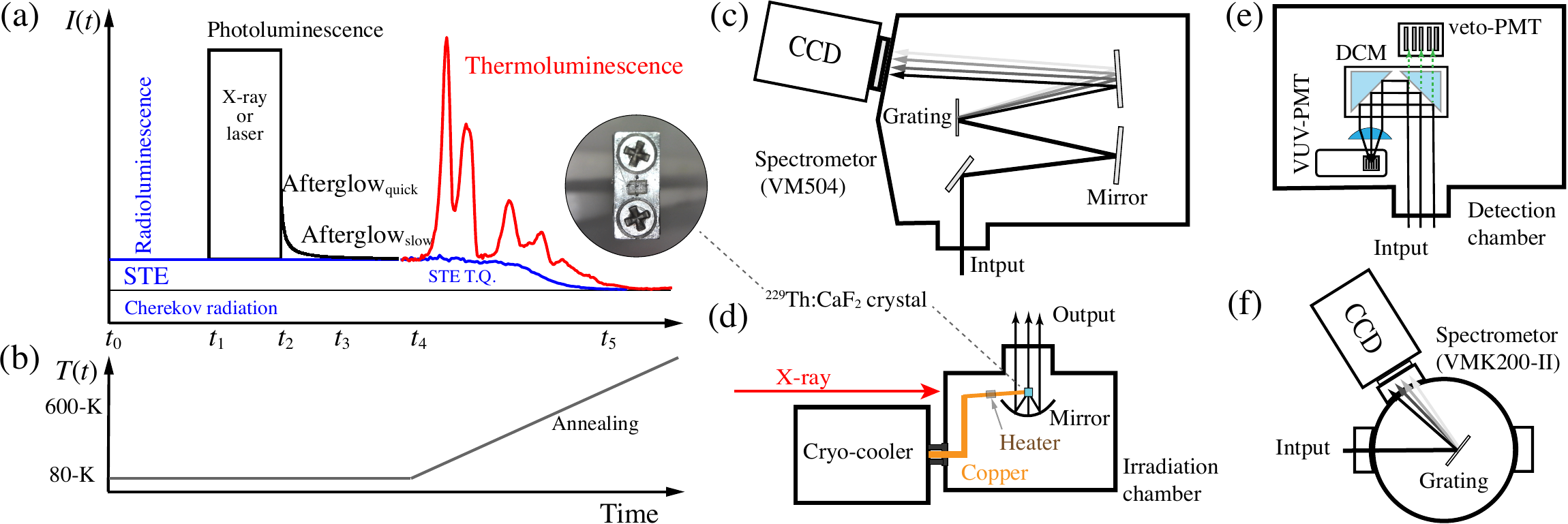}
\caption{(a),(b) Luminescence phenomena in $^{229}$Th:CaF$_2$; (c)--(f) experimental setups used to measure them.
(a) Radioluminescence from energetic $\alpha$ and $\beta$ particles emitted in the decay of $^{229}$Th and its daughters. Photoluminescence occurs when the crystal is exposed to high-energy photons (VUV laser, x-rays, or $\gamma$-rays). Afterglow follows the cessation of ionizing radiation and persists for several minutes. On heating, the STE component of the radioluminescence is thermally quenched and thermoluminescence appears.
The setups in (c)--(f) were developed to characterize this crystal. The crystal is mounted in the irradiation chamber of panel (d), where its temperature is controlled. The chamber output port can be connected to (c) a UV-VIS spectrometer, (e) the detection chamber, or (f) a VUV spectrometer. Details are given in the text.
}
    \label{method}
\end{figure*}

It has recently become clear that the dopant environment is considerably more complex than a single substitutional site. Laser M\"ossbauer spectroscopy has resolved four distinct Th doping sites in CaF$_2$, and it was found that the radiative lifetime is essentially site-independent, whereas the quenching efficiency depends strongly on the site~\cite{hiraki2026laser}. Moreover, continuous-wave absorption spectroscopy resolved inhomogenous broadening of the two dominant centers, one of which exhibits a static crystal-field gradient below \SI{0.1}{\angstrom\tothe{-2}}, two to three orders of magnitude smaller than that of the other center~\cite{morawetz2026continuous}. The structural assignment of the clock-active site is at present still debated~\cite{rehn2026defect}, while x-ray absorption spectroscopy indicates that Th enters as Th$^{4+}$ on a Ca$^{2+}$ site, charge-compensated by fluorine interstitials~\cite{takatori2025xafs}. Fluorine vacancies and their associated in-gap levels have likewise been proposed as states that may interact resonantly with the isomer~\cite{nalikowski2025fluorine}.

The optical response of this defect landscape, however, has not kept pace with these structural advances. The luminescence of Th:CaF$_2$ has so far been studied systematically only in Ref.~\cite{stellmer2015radioluminescence}, at a time when the transition energy was known merely to lie above about 160\,nm; no thermoluminescence study has been reported, and the trap depths that govern the storage and release of charge carriers have never been determined. Because trap-mediated carrier capture is now thought to underlie the quenching mechanisms currently discussed~\cite{guan2025x, elskens2026exploring}, this omission leaves a direct gap between the optical characterization of the material and its use as a clock.

The luminescence phenomena examined in this work are illustrated in Fig.~\ref{method}(a) and (b), and they correspond to the successive stages of a clock interrogation cycle: preparation, excitation, detection, and annealing. Energetic $\alpha$ and $\beta$ particles from the decay of $^{229}$Th and its daughters excite radioluminescence between $t_0$ and $t_1$~\cite{stellmer2015radioluminescence}, which comprises self-trapped exciton (STE) scintillation and Cherenkov radiation~\cite{beeks2023growth}. Between $t_1$ and $t_2$, x-ray or laser irradiation induces intense photoluminescence. After the excitation ceases at $t_2$, an afterglow decays over several minutes~\cite{rao1971afterglow}; for convenience we denote the first $10\,$s as $\mathrm{afterglow_{quick}}$ and the emission at $t_3$ ($5\,$min) as $\mathrm{afterglow_{slow}}$. Once the afterglow has vanished, only radioluminescence remains. The temperature is held constant from $t_0$ to $t_4$; beyond $t_4$, the STE component is thermally quenched [STE T.Q. in Fig.~\ref{method}(a)] and thermoluminescence emerges, its intensity being set by the accumulated color-center density. Finally, near \SI{400}{\celsius} most defects are annealed and the crystal becomes nearly defect-free.

Luminescence offers a window into the defect dynamics of $^{229}$Th:CaF$_2$, which are relevant to both isomer quenching and the optimization of solid-state nuclear clock performance. In this work, we investigate the photoluminescence, radioluminescence, thermoluminescence, and afterglow of $^{229}$Th:CaF$_2$, and use their temperature-dependent behavior to explore the underlying defect processes in the lattice~\cite{hayes2012defects}.

\begin{figure*}[t]
    \centering
    \includegraphics[width=\linewidth]{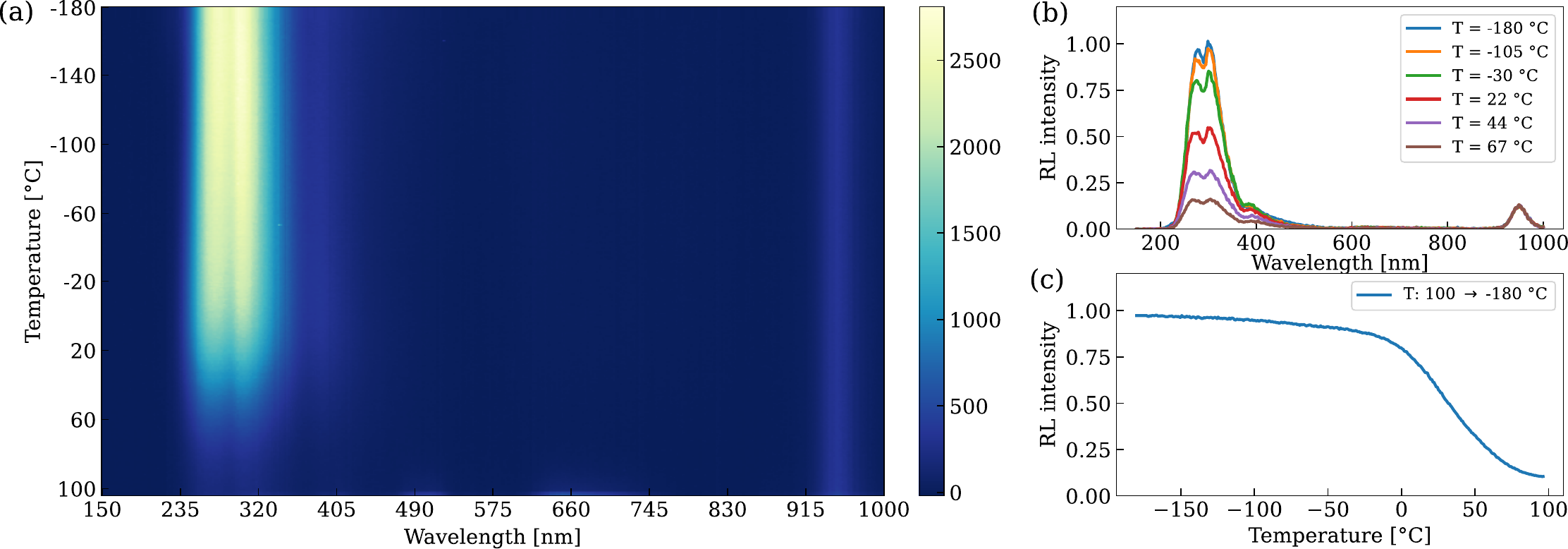}
    \caption{\label{CCD:RL}Thermal quenching of the STE scintillation component of the radioluminescence from the X2 crystal. (a) Spectra measured with a VM504 spectrometer while the crystal was cooled from \SI{100}{\celsius} to \SI{-180}{\celsius} at \C{-0.2}/s, with 5~s CCD exposures. (b) Radioluminescence spectra at selected temperatures. (c) Projection of (a) onto the temperature axis, giving the integrated STE intensity and showing the thermal quenching.}
\end{figure*}

\section{Methods}

The $^{229}$Th:CaF$_2$ crystals used in this study were grown by the TU Wien group, and the samples employed here are listed in Table~\ref{tab:crystals}. In this table, $n(\mathrm{^{229}Th^{4+}})$ denotes the thorium doping concentration, and the ingot names follow the convention established in Refs.~\cite{schreitl2016growth,Beeks:2022hkj}. Small sections were cut from each ingot and used as individual samples; for clarity, each of them is identified throughout by its ingot name.

\begin{table}[b]
    \centering
    \begin{tabular}{cccccc}
    \hline
    \hline
       Crystal ingot  &$n(\mathrm{^{229}Th^{4+}})$ & Activity & Weight & Size & Ref.  \\
    \hline
       V057  & $8\cdot10^{15}$/cm$^3$ & 670\,Bq & 65.5\,mg &19\,mm$^3$& \cite{schreitl2016growth} \\
        X2  & $4\cdot10^{18}$/cm$^3$ & 20000\,Bq & 4.0\,mg &1\,mm$^3$& \cite{Beeks:2022hkj}\\
    \hline
    \hline
    \end{tabular}
    \caption{The $^{229}$Th:CaF$_2$ crystals. The doping concentration is $n(\mathrm{^{229}Th^{4+}})$ and the ingot names follow Refs.~\cite{schreitl2016growth,Beeks:2022hkj}. Sections cut from each ingot were used as samples; their activities and dimensions are listed.}
    \label{tab:crystals}
\end{table}

The experimental arrangement is shown schematically in Fig.~\ref{method}(c)--(f). During the measurements, the crystal was mounted inside the irradiation chamber [Fig.~\ref{method}(d)], where it could be irradiated with x-rays and heated or cooled as required. The photons emitted by the crystal were collected by a parabolic mirror and directed to the flange port, to which either a CCD detector or a PMT-based system was connected, depending on the measurement.

The crystal temperature was controlled by a cryocooler (SunPower, Cryo-GT) combined with ceramic heaters (Thorlabs, HT19R). As shown in Fig.~\ref{method}(d), the 42~K cold head was thermally linked to the stainless-steel crystal holder through a copper rod, which provided efficient cooling, while heaters mounted along the rod supplied heating and temperature stabilization. With this arrangement the temperature could be tuned from \SI{-190}{\celsius} to \SI{300}{\celsius}, as monitored by a Pt100 sensor; further details of the temperature-control system are given in Ref.~\cite{guan2025method}.

To excite the isomeric state and the crystal photoluminescence, we used high-brilliance x-rays from the BL19LXU beamline at SPring-8 (Hyogo, Japan)~\cite{yabashi2001design}. This beamline is equipped with a 27~m undulator and silicon monochromators, which together provide a high-flux ($\sim10^{10}$ photons/s), narrow-bandwidth (30~meV in full width at half maximum) 29.2~keV beam suitable for nuclear resonance scattering~\cite{masuda2019nature,hiraki2024controlling}.

Luminescence spectra were recorded with the irradiation chamber connected to one of two CCD-based spectrometers, as shown in Fig.~\ref{method}(c) and (f). Depending on the wavelength range of interest, we used either a Czerny--Turner spectrometer (Princeton Instruments, VM504)~\cite{vm504} or a Seya--Namioka spectrometer (Vacuum Optics, VMK200-II)~\cite{vmk200ii}. The VM504 was employed for the ultraviolet to near-infrared range, whereas the VMK200-II was reserved for VUV measurements of $^{229\mathrm{m}}$Th signals and of the transmission and reflectance of optical components. Either spectrometer accepts a CCD detector (Teledyne, PIXIS-XO:100BSO)~\cite{pixis-xo}. All measurements reported here used a 600~g/mm grating blazed near 300~nm together with a 2~mm entrance slit, and the wavelength scale was calibrated with mercury, cadmium, and argon lamps.

The detection chamber shown in Fig.~\ref{method}(e) is designed to detect VUV photons emitted by $^{229\mathrm{m}}$Th. Inside it, a solar-blind photomultiplier (VUV-PMT; Hamamatsu R10454) detects the isomer signal, while a UV-sensitive photomultiplier (veto-PMT; Hamamatsu R11265-203) monitors the radioluminescence background. Dichroic mirrors reflecting near 150~nm suppress background photons, and an MgF$_2$ lens focuses the filtered light onto the VUV-PMT. The signals from both tubes were amplified and digitized as waveforms by a high-speed oscilloscope (National Instruments, PXIe-5162). Because the VUV-PMT responds mainly between 130 and 160~nm and the veto-PMT between 200 and 550~nm, the two detectors are respectively sensitive to the Cherenkov and STE components of the radioluminescence.

All measurements were performed under vacuum, as is required both for VUV detection and for low-temperature operation. After optimization, the pressure in the irradiation and detection chamber reached the $10^{-4}$~Pa level; when the cryocooler was running, the cold head and copper rod acted as cold traps and improved it further to $10^{-5}$~Pa~\cite{guan2025method}.

 \begin{figure*}[t]
    \centering
    \includegraphics[width=\linewidth]{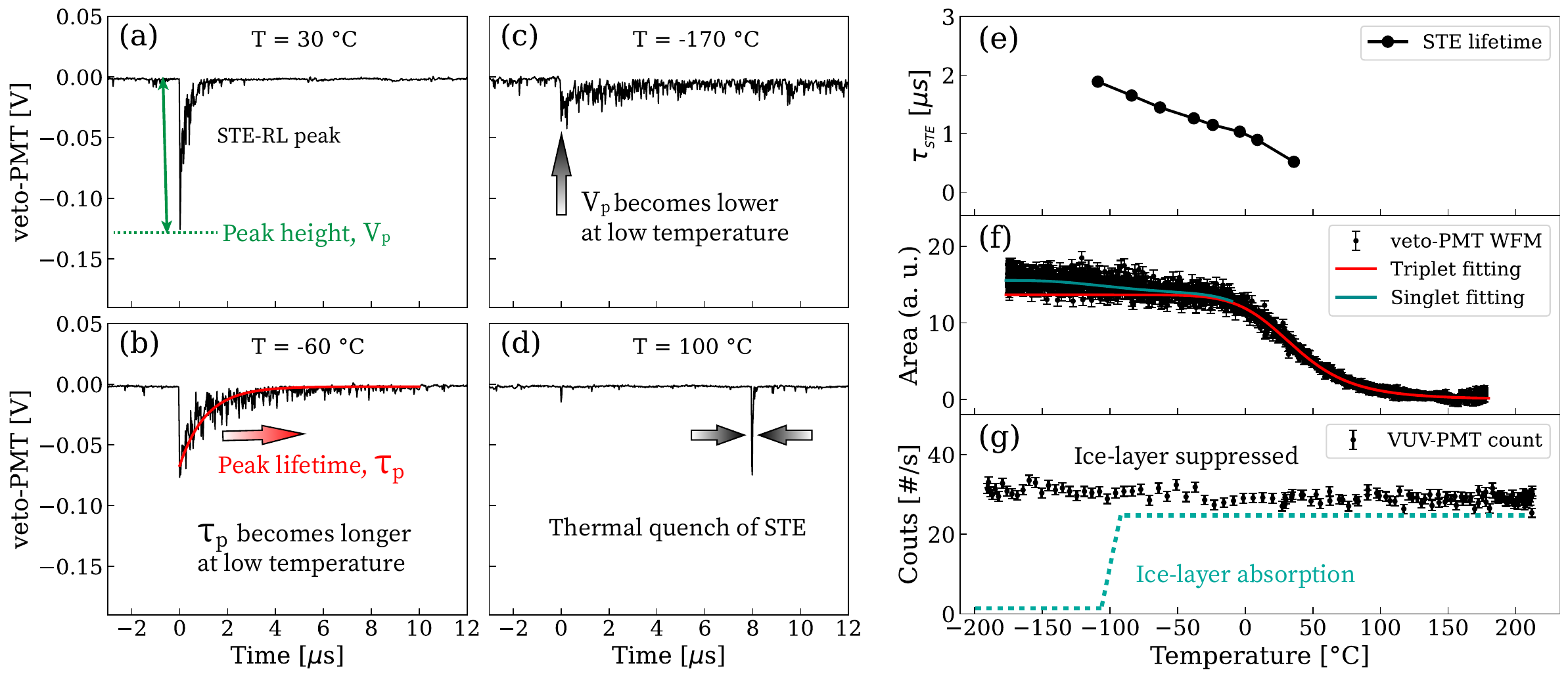}
    \caption{\label{veto-PMT-RL} The radioluminescence measured with the PMT-based setup. (a)--(d) Waveforms of the radioluminescence signal measured by the veto-PMT at different crystal temperatures. Each waveform consists of a light burst caused by the emission of an $\alpha$ or $\beta$ particle, followed by an exponentially decaying tail. (a) Waveform recorded at \SI{30}{\celsius}, defining the peak height $V_p$ and the decay constant $\tau_p$; the tail is fitted with Eq.~\eqref{equation_tau}. (b) At \SI{-60}{\celsius}, the decay constant $\tau_p$ becomes longer. (c) At \SI{-170}{\celsius}, the peak height $V_p$ becomes lower. (d) At \SI{100}{\celsius}, the STE scintillation is thermally quenched and the waveform collapses into a short-tailed spike. (e) The decay constant $\tau_p$ of the STE light burst, fitted from the veto-PMT waveforms, as a function of temperature. (f) Integrated veto-PMT waveform area as a function of temperature; the area is proportional to the radioluminescence light yield. The black points are the data. The red curve shows the fit including only the triplet STE component, while the cyan curve shows the total fit including both the triplet and singlet components. The dashed green curve indicates the singlet contribution to the total fit. (g) VUV-PMT counting rate as a function of temperature. The cyan dotted curve shows the counting rate when an ice layer is allowed to form on the crystal surface at low temperature, which absorbs the Cherenkov photons; the black points show the rate measured with the cryogenic copper shield installed, which suppresses the ice-layer formation.}
    \end{figure*}

\section{Radioluminescence \label{label_radioluminescence}}

Radioluminescence originates from photons emitted within the crystal itself, and it constitutes the dominant optical background for clock operation. We therefore recorded its spectrum from the ultraviolet to the infrared with the VM504 spectrometer, as shown in Fig.~\ref{CCD:RL}. Since the 600~g/mm grating projects only a 100~nm window onto the CCD, the full spectrum had to be assembled from nine successive measurements, the grating being rotated by 100~nm between them. To examine the thermal quenching of the STE component, the crystal was cooled from \SI{100}{\celsius} to \SI{-180}{\celsius} over the course of about 30 minutes, during which spectra were taken with 5~s exposures; the nine segments were then stitched together by matching the data at identical temperatures.

\begin{figure*}[t]
    \centering 
    \includegraphics[width=0.98\linewidth]{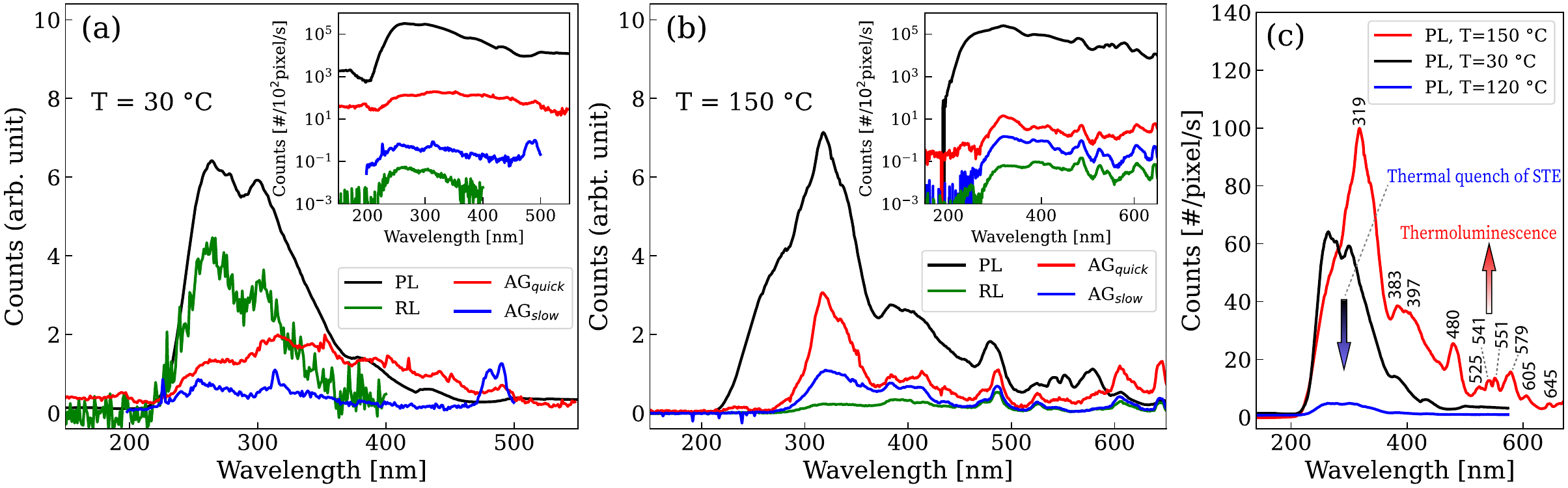}
    \caption{Luminescence of the V057 crystal measured at SPring-8. (a) July 2019, at \SI{30}{\celsius}. (b) October 2019, at \SI{150}{\celsius}. (c) Both runs, at \SI{30}{\celsius}, \SI{120}{\celsius}, and \SI{150}{\celsius}.}
    \label{photoluminescence}
\end{figure*}

The resulting spectrum [Fig.~\ref{CCD:RL}(a)] exhibits two bright bands near 275 and 310~nm, in agreement with earlier measurements~\cite{stellmer2015radioluminescence}, together with a weaker band near 400~nm. Figure~\ref{CCD:RL}(b) shows spectra recorded at several temperatures, in which the STE scintillation band, a sub-band peaking at 400~nm, and a peak near 950~nm are visible; the last of these is insensitive to temperature and its origin is at present unknown. The radioluminescence is weak at elevated temperatures and becomes progressively stronger as the crystal is cooled. Figure~\ref{veto-PMT-RL}(f) shows the temperature-dependent STE yield, which clearly demonstrates the thermal quenching~\cite{mikhailik2006scintillation}.

Radioluminescence is generated by energetic $\alpha$ and $\beta$ particles emitted in the $^{229}$Th decay chain. Both particle types excite scintillation light bursts, i.e., STE luminescence, and the faster $\beta$ particles additionally produce Cherenkov radiation~\cite{gong2024feasibility}. These STE bursts are registered by the veto-PMT as individual peaks in the oscilloscope waveform [Fig.~\ref{veto-PMT-RL}(a)--(d)]. The area of a peak, which measures the number of photons contained in a burst, is determined by the peak height $V_p$ and by the decay time $\tau_p$ of its tail. For brevity, we refer to these bursts throughout as STE-RL.

Large peaks are $\alpha$ events and smaller ones $\beta$ events. The tail is fitted with

\begin{equation}
    V(t) = V_p \exp\left(-\frac{t - t_0}{\tau_p}\right), \label{equation_tau}
\end{equation}

where $t_0$ denotes the position of the peak, as illustrated in Fig.~\ref{veto-PMT-RL}(a). At room temperature this fit yields a decay constant $\tau_p \approx 1$~$\mu$s.

Cooling the crystal alters the waveform markedly [Figs.~\ref{veto-PMT-RL}(b) and \ref{veto-PMT-RL}(c)]: the peak height decreases while the tail becomes longer. Heating, by contrast, suppresses the signal, and at \SI{100}{\celsius} the waveform is reduced to a nearly vertical spike [Fig.~\ref{veto-PMT-RL}(d)]. As the fitted values in Fig.~\ref{veto-PMT-RL}(e) show, $\tau_p$ increases as the temperature decreases, so that the scintillation lifetime becomes longer when the crystal is cold.

The integral of the veto-PMT waveform provides a second, independent measure of the temperature dependence of the STE light yield. During the cooling experiment, this yield was found to increase at lower temperatures and to fall to nearly zero above \SI{100}{\celsius}. In a dedicated measurement, the oscilloscope was therefore configured to record 1~ms-long veto-PMT waveforms at a 100~Hz acquisition rate while the X2 crystal was gradually cooled from \SI{170}{\celsius} to \SI{-190}{\celsius}. The integrated waveform areas, which represent the total STE light yield, are shown in Fig.~\ref{veto-PMT-RL}(f); their overall trend agrees well both with the thermal quenching observed in the spectrometer-based radioluminescence measurements and with the behavior reported in Refs.~\cite{rodnyi2020physical, Beeks:2022hkj}. To describe this temperature dependence quantitatively, we fit the data with the standard thermal quenching relation for STEs in alkaline earth fluorides~\cite{rodnyi2020physical},

\begin{equation}
p = \frac{p_r}{p_r + p_{nr}} = \frac{1}{1 + C_q \exp\left(-\frac{E_q}{kT}\right)}, \label{equation_quench}
\end{equation}

where $p$ is the probability of radiative decay, $p_r$ and $p_{nr}$ are the radiative and non-radiative decay probabilities of an STE, and $C_q$ and $E_q$ are the quenching constant and the quenching activation energy, respectively. It is well established that the STE in CaF$_2$ possesses \textit{triplet} and \textit{singlet} components~\cite{williams1976time, williams1990self, rodnyi2020physical}. Fitting the triplet component alone, shown as the red curve, yields $C_q^{\mathrm{triplet}} = (1.39 \pm 0.017) \times 10^{6}$ and $E_q^{\mathrm{triplet}} = 0.3767 \pm 0.005$\,eV. The deviation between this fit and the data at low temperatures is ascribed to the singlet component, which is described by the offset fit shown as the cyan curve, giving $C_q^{\mathrm{singlet}} = 775 \pm 236$ and $E_q^{\mathrm{singlet}} = 0.102 \pm 0.006$\,eV. The singlet STE therefore has a lower quenching activation energy and a smaller amplitude than the triplet, the intensity ratio being approximately 1:7.

Cherenkov radiation forms the second component of the radioluminescence~\cite{stellmer2016feasibility}. It dominates the region between 120~nm, the cutoff imposed by the bandgap, and 200~nm, and its intensity falls approximately as $1/\lambda^3$~\cite{stellmer2016feasibility, beeks2023growth}. In marked contrast to STE-RL, this component is nearly independent of temperature. Below \SI{-100}{\celsius}, however, residual gas condenses on the crystal and forms an ice layer, which absorbs Cherenkov photons and, more importantly, the 148~nm isomer photons~\cite{tiedau2024laser}.

In the detection chamber [Fig.~\ref{method}(e)], the VUV-PMT is likewise sensitive to Cherenkov photons near 150~nm. When the crystal was cooled for the isomer measurements, an ice layer formed and absorbed them, as indicated by the dashed line in Fig.~\ref{veto-PMT-RL}(g). To prevent this, we enclosed the crystal in a cryogenic copper shield fitted with an MgF$_2$ window~\cite{guan2025method}. Once the shield had been installed, the VUV-PMT rate remained stable at approximately 30~Hz while the crystal was cooled from \SI{220}{\celsius} to \SI{-200}{\celsius} [Fig.~\ref{veto-PMT-RL}(g)]. This confirms both that ice formation was suppressed and that the Cherenkov component is indeed temperature-independent, in contrast to the strongly quenched STE radioluminescence.

\section{Photoluminescence \label{label_photoluminescence}}

Photoluminescence is induced when the crystal is exposed to x-ray or laser irradiation. Pure CaF$_2$ scintillates through the STE mechanism and exhibits the same 280~nm band as the radioluminescence~\cite{rodnyi2020physical}. Activators, which are mainly trivalent lanthanides, introduce additional luminescent centers whose bands originate from $5d\rightarrow 4f$ transitions of the excited $4f$ electron~\cite{yanagida2010growth, Kolaly1981effect}. The Th$^{4+}$ dopant, by contrast, has a closed electronic shell and, as will be shown below, produces no clear emission bands of its own.

We measured the photoluminescence spectra of the V057 crystal at \SI{30}{\celsius}, \SI{120}{\celsius}, and \SI{150}{\celsius} under 29.2\,keV synchrotron irradiation with a photon flux  of $\sim10^{11}$ photons/s for 60\,s, subsequently recorded the afterglow that follows the termination of the beam (Fig.~\ref{photoluminescence}). In these figures the red (``quick'') and blue (``slow'') curves show the afterglow measured starting within 1s and 5\,min after the beam was turned off, respectively, while the green curves show the radioluminescence recorded for comparison. Because the three signals differ greatly in intensity, the CCD pixels were binned differently for each exposure: the photoluminescence was limited to 20~ms to avoid saturating or damaging the sensor, the much weaker radioluminescence required 500~s, and the afterglow was recorded with exposures of 10~s (afterglow$_\mathrm{quick}$) and 250~s (afterglow$_\mathrm{slow}$).

At room temperature the radioluminescence and photoluminescence spectra are very similar, and both are dominated by the 280~nm STE band [Fig.~\ref{photoluminescence}(a)]. Once the excitation is switched off, however, the afterglow reveals several red-shifted bands: emission between 300 and 400~nm dominates during the first 10~s, whereas after 5~min the distribution has shifted and a doublet structure appears near 480--490~nm.

The inset of Fig.~\ref{photoluminescence}(a) normalizes the counting rates to counts per second per $10 \times 10$ pixel area, which allows the three signals to be compared directly. The photoluminescence is about $\mathcal{O}(10^3)$ times stronger than afterglow$_\mathrm{quick}$, and the afterglow weakens by roughly a factor of 50 within 5~min. Radioluminescence exhibits the lowest rate of the three, approximately an order of magnitude below afterglow$_\mathrm{slow}$.

At \SI{150}{\celsius} the spectra change markedly [Fig.~\ref{photoluminescence}(b)]. Under these conditions the photoluminescence and afterglow acquire nearly the same spectral distribution and differ only in intensity, because thermal quenching suppresses the 280~nm band so that other mechanisms dominate, giving rise to peaks near 320, 400, and 490~nm.

 \begin{figure*}[t]
    \centering 
    \includegraphics[width=\linewidth]{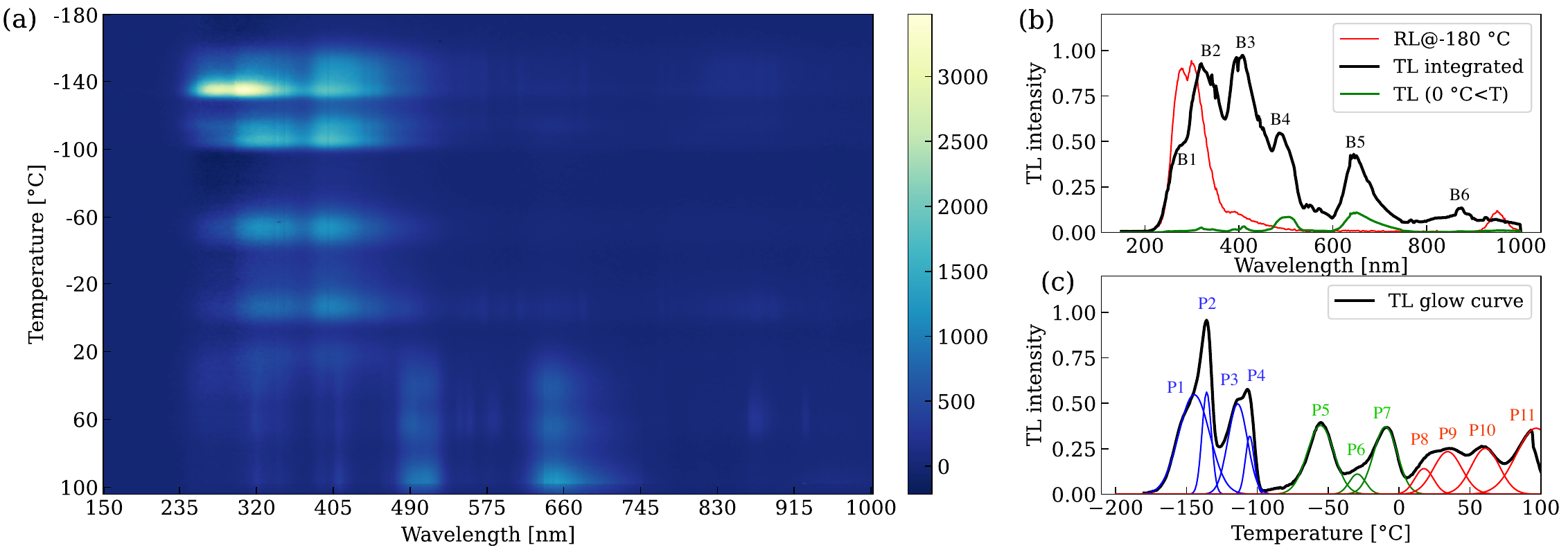}
    \caption{Thermoluminescence of the X2 crystal measured with the VM504 spectrometer. The crystal was held at \SI{-180}{\celsius} for 16~h to accumulate color centers and then heated to \SI{100}{\celsius} at \SI{0.2}{\celsius}/s. (a) Spectra taken with 5~s exposures; the full range is assembled from nine 100~nm windows. (b) Projection of (a) onto the wavelength axis, showing the emission bands labeled B. The red curve is the radioluminescence at \SI{-180}{\celsius} and the green curve the thermoluminescence intensity integrated above \SI{0}{\celsius}. (c) Projection onto the temperature axis, i.e., the glow curve, with peaks labeled P. For the last peak P11 the subtraction of the radioluminescence is uncertain because charge carriers remained trapped at the start of cooling; P11 is fitted at \SI{96.5}{\celsius}, close to the peak reported in Ref.~\cite{ratnam1966thermoluminescence}.}
    \label{2024.10:CCD-TL}
\end{figure*}

The inset of Fig.~\ref{photoluminescence}(b) shows the temporal evolution of the luminescence during and after the irradiation. Here the afterglow decays much faster than at room temperature: once the beam is switched off, the photoluminescence falls by $\mathcal{O}(10^4)$ within 10~s, down to the level of afterglow$_\mathrm{quick}$. A comparison of the insets of Figs.~\ref{photoluminescence}(a) and \ref{photoluminescence}(b) further shows that the intensity differences among afterglow$_\mathrm{quick}$, afterglow$_\mathrm{slow}$, and radioluminescence become much smaller at elevated temperature, which indicates that the afterglow fades more rapidly when the crystal is hot.

Figure~\ref{photoluminescence}(c) compares the photoluminescence spectra recorded at the three temperatures on a counts-per-pixel-per-second scale. It is evident that temperature affects both the intensity and the spectral distribution. At room temperature the spectrum consists of a moderate STE band; at \SI{120}{\celsius} this band is quenched; and at \SI{150}{\celsius} the photoluminescence re-emerges with greater intensity and is shifted towards longer wavelengths, an effect that we attribute to the thermal activation of color-center recombination involving deeper traps. As will become clear in the following section, this thermally activated emission is in fact the thermoluminescence.

\begin{figure*}
    \centering
    \includegraphics[width=\linewidth]{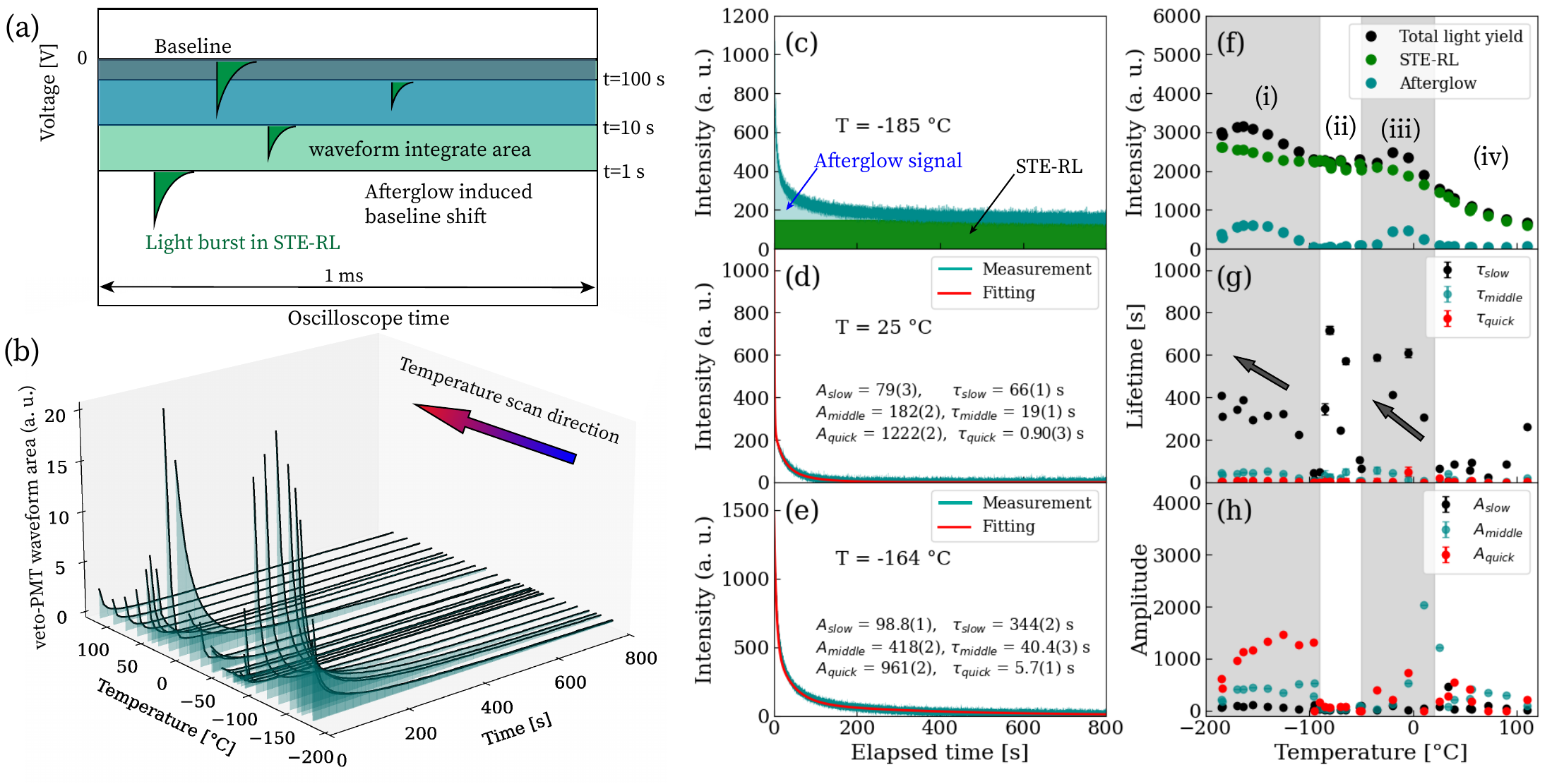}
    \caption{Measurement of the afterglow of the $^{229}$Th:CaF$_2$ crystal. The crystal was irradiated with the beamline x-ray for 1000\,s, after which the afterglow was recorded for 1800\,s. The temperature was varied from \SI{-190}{\celsius} to \SI{110}{\celsius} in steps of \SI{10}{\celsius}. (a) Schematic of the veto-PMT waveform at three successive times after the x-ray beam was switched off. The afterglow shifts the waveform baseline downward, while the STE-RL light bursts appear as superimposed peaks; integrating the waveform over its 1\,ms duration therefore yields the sum of both contributions. (b) Three-dimensional representation of the integrated veto-PMT waveform area as a function of elapsed time and crystal temperature. (c) Afterglow decay recorded at \SI{-185}{\celsius}; the green area represents the constant STE-RL background, and the cyan area the pure afterglow after subtraction of this background. (d),(e) Examples of the afterglow decay fitted with three exponential components at \SI{25}{\celsius} and \SI{-164}{\celsius}, respectively; the fitted amplitudes $A_i$ and lifetimes $\tau_i$ are annotated. (f) Temperature dependence of the total light yield (black), the constant STE-RL background (green), and the pure afterglow (cyan). The grey shaded bands labeled (i)--(iv) indicate the temperature regions discussed in the text. (g),(h) Fitted lifetimes and amplitudes of the three afterglow components, respectively.}
    \label{afterglow}
\end{figure*}


\section{Thermoluminescence \label{label_thermoluminescence}}

CaF$_2$ is a well-known thermoluminescent material and has been studied extensively for dosimetry applications~\cite{kafadar2009effects, kafadar2011thermal}. Its thermoluminescence is influenced by impurities, crystal defects, and thermal history: when rare-earth activators are present, the emission is determined largely by the dopant ions~\cite{merz1967charge1, merz1967charge2, de2024thermoluminescence, Kolaly1981effect}, whereas in undoped or weakly doped crystals it is governed instead by intrinsic recombination at native defects and charge traps~\cite{atobe1979thermoluminescence, ratnam1966thermoluminescence}.

In our measurements, self-irradiation from the thorium decay chain continuously excites radioluminescence and simultaneously creates defects in the lattice. When the crystal is subsequently heated, these radiation-induced traps are released and thermoluminescence is emitted.

The thermoluminescence spectra were measured using the VM504 spectrometer. The X2 crystal was first held at \SI{-180}{\celsius} for 16~h, so that self-irradiation could accumulate color centers, and was then heated at a constant rate of \SI{0.2}{\celsius}/s. Upon reaching \SI{100}{\celsius} the heating was paused for 10~min in order to bleach the carriers trapped below that temperature, after which the crystal was cooled back to \SI{-180}{\celsius} at \SI{-0.2}{\celsius}/s and held there for a further 16~h before the cycle was repeated. A custom PID program controlled the temperature throughout. As described above, the 600~g/mm grating covers only about 100~nm per measurement, so nine consecutive measurements at different grating positions were required to span the ultraviolet to the near-infrared range.

Each measurement covered the whole warming--bleaching--cooling cycle. The heating phase contains thermoluminescence and radioluminescence, whereas the cooling phase contains radioluminescence alone; subtracting the latter (Fig.~\ref{CCD:RL}) from the former gives the net thermoluminescence. The full spectrum was then assembled from the nine windows (Fig.~\ref{2024.10:CCD-TL}).

\begin{table}[b]
    \centering
    \begin{tabular}{ccclcc}
    \hline
    \hline
       & Wavelength & FWHM &TL peaks & Note& Ref. \\
    \hline
     B1   & 263(1) & 16 & P1-P4& 1st peak in RL & \cite{atobe1979thermoluminescence} \\
     B2   & 323(1) & 29 & P1-P7& 2nd peak in RL& \cite{atobe1979thermoluminescence, ratnam1966thermoluminescence} \\
     B3   & 407(1) & 30 & P1-P7& 3rd peak in RL \\
     B4   & 490(1) & 29 & P8-P11 & &\cite{atobe1979thermoluminescence}\\
     B5   & 652(1) & 29 & P8-P11& &\cite{ratnam1966thermoluminescence} \\
     B6   & 854(6) & - & P1-P10 & & \\
    \hline
    \hline
    \end{tabular}
    \caption{Emission bands in the thermoluminescence measured with the CCD spectrometer. Band B6 contains many weak emission centers whose structure is not resolved here.}
    \label{bands}
\end{table}
\begin{table}[b]
    \centering
    \begin{tabular}{lccclll}
    \hline
    \hline
         & \multicolumn{2}{l}{Peak temperature}& FWHM & Activation &  Emission & Ref. \\
         \cline{2-3}
         & [Kelvin] & [Celsius]&[K/$^\circ$C] &energy [eV] &  Band\\
    \hline
       P1  & 128.4(3)& -144.8(3)&25.7 & 0.115(2) & B1-B3 & \cite{atobe1979thermoluminescence}\\
       P2  & 137.3(1)& -135.9(1)&7.7& & B1-B3\\
       P3  & 159.1(3)& -114.1(3)& 16.1& 0.125(2) & B1-B3 & \\
       P4  & 167.6(1)& -105.6(1)& 7.2 & & B1-B3\\
       \hline
       P5  & 217.6(2)& -55.6(2)& 22.2& 0.372(3) & B2, B3 & \cite{atobe1979thermoluminescence}\\
       P6  & 243.6(3)& -29.6(3) & 13.3 & &B2, B3\\
       P7  & 263.4(7)& -9.8(7)& 20.6 & &B2, B3\\
       \hline
       P8  & 297.0(3)& 23.8(3) & 28.3 & 0.503(2)  & B4, B5\\
       P9  & 312.8(1)& 39.6(1) &  15.9 &  & B4, B5\\
       P10  & 332.1(1)& 58.9(1) &22.2 &  & B4, B5\\
       P11  & 369.7(4)& 96.5(4) &32.3 & 0.535(1) & B4, B5 & \cite{ratnam1966thermoluminescence}\\
    \hline
    \hline
    \end{tabular}
    \caption{Thermoluminescence peaks in the glow curve, measured with the CCD spectrometer. Peak positions and widths are from a multi-Gaussian fit of the glow curve [Fig.~\ref{2024.10:CCD-TL}(c)]. The activation energy $E$ of each peak is obtained by fitting its low-temperature leading edge with a function proportional to $\exp(-E/kT)$, with $k$ the Boltzmann constant and $T$ the absolute temperature.}
    \label{peaks}
\end{table}

The thermoluminescence spectra of Fig.~\ref{2024.10:CCD-TL}(a) contain several distinct emission bands together with isolated glow peaks; their projections onto the wavelength and temperature axes are presented in Fig.~\ref{2024.10:CCD-TL}(b) and \ref{2024.10:CCD-TL}(c), respectively. Six emission bands, labeled B, are visible in Fig.~\ref{2024.10:CCD-TL}(b), where the STE-RL spectrum is overlaid in red for comparison. The centers of B1 and B2 coincide with the STE-RL doublet, being shifted slightly to the blue and to the red respectively, while B3 aligns with the weak STE-RL peak near 400~nm. Bands B4 and B5, centered near 490 and 650~nm, become prominent only above room temperature, as highlighted by the green curve. Band B6 consists of many weak peaks and appears to be centered near 850~nm. Table~\ref{bands} compares these bands with those reported in earlier work.

The glow curve [Fig.~\ref{2024.10:CCD-TL}(c)] contains the peaks labeled P, which fall into three groups according to their temperature. The first group spans \SI{-180}{\celsius} to \SI{-100}{\celsius} and contains P1--P4, which were fitted with multi-Gaussian functions (blue curves); the corresponding spectra show that these peaks arise mainly from bands B1--B3. Above \SI{-100}{\celsius} the signal decreases until P5 emerges near \SI{-60}{\celsius}, marking the beginning of the second group, which extends from \SI{-80}{\celsius} to \SI{10}{\celsius} and contains P5--P7, again dominated by B1--B3. The third group, from \SI{10}{\celsius} to \SI{100}{\celsius}, contains the overlapping peaks P8--P11, which are associated with bands B4 and B5. In addition to these main features, weaker isolated peaks appear across the visible to near-infrared range. Table~\ref{peaks} lists the fitted peaks and their correspondence with earlier observations.

These observations indicate that the thermoluminescence of $^{229}$Th:CaF$_2$ resembles that of pure CaF$_2$~\cite{atobe1979thermoluminescence} rather than that of lanthanide-doped CaF$_2$~\cite{merz1967charge1, merz1967charge2, de2024thermoluminescence}, so Th$^{4+}$ acts as an optical activator only weakly, if at all. The weak isolated peaks among the main bands may instead arise from thorium ions or their decay daughters. As discussed below, thorium introduces in-gap levels that can trap electrons, and thermally activated holes migrating to thorium sites would contribute to the emission. This may explain band B6 [Fig.~\ref{2024.10:CCD-TL} and Table~\ref{bands}]. Resolving the interplay between the dopant and the host requires further study.

\section{Afterglow \label{label_afterglow}}
Long-lived fluorescence, commonly termed afterglow, persists for several minutes in fluoride crystals after the ionizing radiation has ceased. It arises from the recombination of spatially separated F and H centers: whereas the STE involves nearest-neighbor F/H pairs and decays within microseconds, the afterglow requires a color center to hop across several lattice sites before it encounters its counterpart. The afterglow thus originates from well-separated F and H centers and persists over far longer timescales than STE scintillation.

The afterglow was measured at SPring-8. The X2 crystal was first irradiated with a 29.2~keV x-ray beam at $10^{11}$ photons/s for 1000~s, after which the beam was switched off and the veto-PMT recorded the afterglow for 1800~s. The oscilloscope was triggered at a fixed rate of 100~Hz and stored 1~ms waveforms, which were subsequently integrated to give the afterglow intensity. As shown in Fig.~\ref{afterglow}(a), the afterglow shifted the waveform baseline downward immediately after the beam was switched off, with the STE-RL peaks superimposed on this shifted baseline. Because the baseline returned towards 0~V as the afterglow decayed, each integrated waveform contains contributions from both processes.

After each 1000~s irradiation and the subsequent 1800~s detection period, the crystal was set to the next temperature, the range extending from \SI{-190}{\celsius} to \SI{110}{\celsius} in steps of \SI{10}{\celsius}. Figure~\ref{afterglow}(b) shows the afterglow intensity as a function of time, averaged over 10~s intervals. Since a constant STE-RL component was present throughout the measurements, it was subtracted in order to isolate the pure afterglow; the result obtained at \SI{-185}{\celsius} is shown in Fig.~\ref{afterglow}(c), where the STE-RL background is indicated in green and the afterglow in cyan.

\begin{figure*}[t]
    \includegraphics[width=\textwidth]{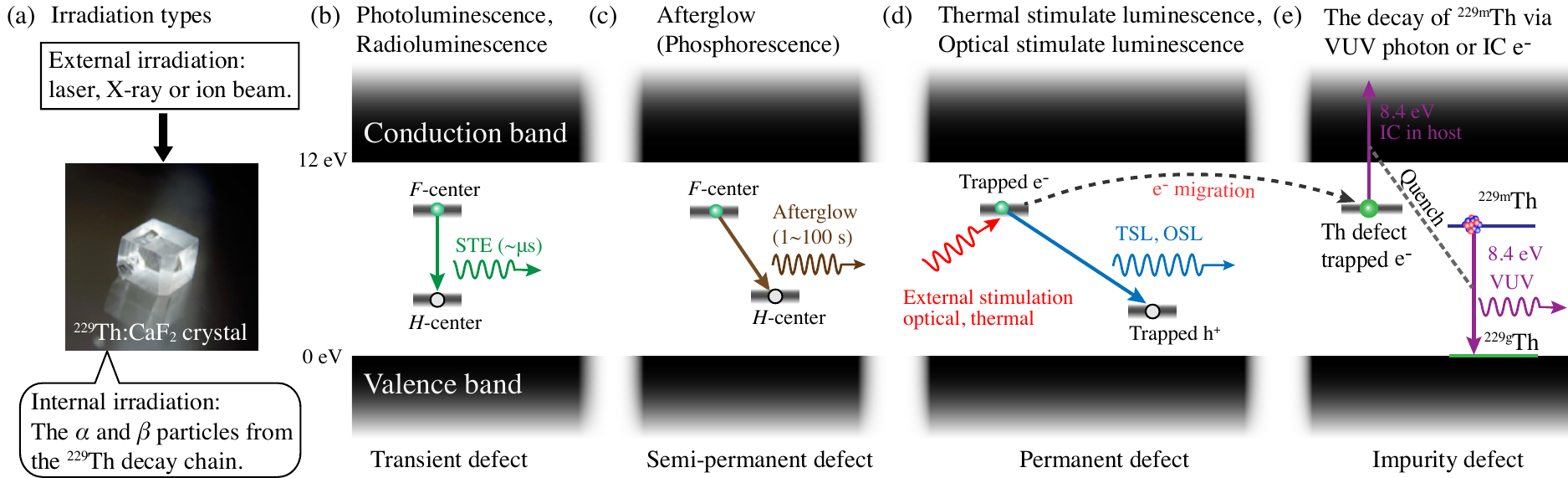}
\caption{Origins and consequences of the luminescence discussed in this work. (a) External and internal radiation both induce luminescence in $^{229}$Th:CaF$_2$.
(b) Recombination of nearest-neighbor F and H centers gives prompt STE scintillation, seen in photoluminescence and radioluminescence alike.
(c) Recombination of closely spaced F and H centers produces the afterglow.
(d) Recombination of widely separated trapped electrons and holes gives thermally or optically stimulated luminescence: on stimulation the carriers escape, migrate, and recombine.
(e) Doping with $^{229}$Th$^{4+}$ introduces in-gap levels that trap migrating electrons, which open a non-radiative decay channel of the isomer, known as \textit{isomer quenching}.
\label{defects}
}
\end{figure*}

Once the STE-RL and afterglow contributions had been separated at each temperature, the net decay curves of the afterglow were obtained, two examples of which are shown in Figs.~\ref{afterglow}(d) and \ref{afterglow}(e). Figure~\ref{afterglow}(f) summarizes the temperature dependence of the total, STE-RL, and afterglow light yields. As can be seen, the STE-RL component (green points) follows the thermal quenching described in the preceding sections, whereas the afterglow component (cyan points) exhibits a temperature dependence whose amplitude is comparable to that observed in the thermoluminescence measurements.

The afterglow decay curves were fitted with a sum of three exponentials, i.e.,

\begin{equation}
    L_{\mathrm{afterglow}}(t) = \sum_i A_i \exp\left(-\frac{t}{\tau_i}\right),
    \label{3exp}
\end{equation}

where $A_i$ and $\tau_i$ denote the amplitude and the lifetime of component $i$; these three components are labeled \textit{quick}, \textit{middle}, and \textit{slow} in order of increasing lifetime. Figures~\ref{afterglow}(d) and \ref{afterglow}(e) show two examples of such fits, obtained at \SI{25}{\celsius} and \SI{-164}{\celsius}. The integrated afterglow yield differs greatly between these two temperatures, and every component becomes longer-lived as the crystal is cooled.

Figures~\ref{afterglow}(g) and \ref{afterglow}(h) summarize the fitted lifetimes and amplitudes for all temperatures. The three components have lifetimes near 1, 10, and 100~s, respectively, and within each of the temperature groups defined in the thermoluminescence section the lifetimes increase as the temperature decreases, as indicated by the black arrows. This trend is most pronounced for the slow component between \SI{-190}{\celsius} and \SI{-100}{\celsius} [Fig.~\ref{afterglow}(g)]. Moreover, the amplitudes of the individual components follow the same temperature dependence as the thermoluminescence intensity in the corresponding range, which again points to a common origin.

\section{Discussion}
\subsection{Defect and luminescences}
The luminescence measured here closely resembles that of pure CaF$_2$~\cite{mikhailik2006scintillation, atobe1979thermoluminescence, rao1971afterglow}. This similarity indicates that thorium does not act as an optical activator, in contrast to the rare-earth-doped CaF$_2$ scintillators, in which the emission is dominated by the dopant ions.

To interpret the observed luminescence systematically, it is useful to distinguish two kinds of ionizing radiation, as illustrated in Fig.~\ref{defects}(a). The first is internal, originating from the $\alpha$ and $\beta$ particles emitted in the decay chain of $^{229}$Th, and the second is external, such as the x-rays or laser light used to excite the nuclei. Whereas the external radiation can be switched on and off at will, the internal radiation gives rise to a continuous radioluminescence that cannot be interrupted.

The luminescence can furthermore be described in terms of three classes of lattice defect~\cite{rodnyi2020physical, hayes2012defects}, which are summarized in Figs.~\ref{defects}(b)--(d). In these illustrations the upper and lower bands represent the conduction and valence bands, separated by an energy gap of 12~eV in pure CaF$_2$, while the defects are depicted as electron traps, such as F centers located near the bottom of the conduction band, and hole traps, such as H centers situated near the top of the valence band. The horizontal spacing between the defect symbols indicates the spatial separation of the defects in the lattice.

The transient defects [Fig.~\ref{defects}(b)] are nearest-neighbor F--H pairs, i.e., STEs. They recombine within microseconds and give rise to the dominant component of both the photoluminescence and the radioluminescence.

The semi-permanent defects [Fig.~\ref{defects}(c)] are F and H centers separated by several lattice sites. When the irradiation is switched off, the nearest-neighbor pairs recombine promptly, but the lattice remains phonon-rich for a short time, so that the remaining separated centers can still acquire enough energy to migrate. The recombination of a migrating electron with a hole then emits a photon on a much longer timescale, and this delayed emission constitutes the afterglow. As the lattice gradually cools, migration slows down, the afterglow fades, and the surviving centers become immobilized as permanent defects.

The permanent defects [Fig.~\ref{defects}(d)] are widely separated and immobile color centers that remain stable at a given temperature. Thermal or optical stimulation releases the trapped carriers, which migrate through the lattice and recombine with carriers of the opposite type, giving rise to thermally stimulated luminescence (TSL) or optically stimulated luminescence (OSL) depending on the nature of the stimulation. Such irradiation--stimulation--emission cycles are widely exploited in dosimetry~\cite{yukihara2011optically, kafadar2011thermal}.

\subsection{Defect and isomer quenching}
As Th:CaF$_2$ is one of the leading platforms for a solid-state clock, understanding the coupling between the nucleus and its host is of particular interest for improving and optimizing future clock operation. When radiation is tuned to the $^{229}$Th resonance, the nucleus is excited to its isomeric state~\cite{hiraki2024controlling, tiedau2024laser, zhang2024frequency}. As illustrated in Fig.~\ref{defects}(e), the isomer energy of 8.3557\,eV lies well below the bandgap of CaF$_2$, so that the VUV photons emitted during the decay can escape the host; moreover, because the dopant is tetravalent, the free-electron internal conversion channel that dominates in neutral atoms is energetically closed. External stimulation nevertheless opens additional decay channels that shorten the isomeric lifetime, an effect known as \textit{isomer quenching}~\cite{hiraki2024controlling, schaden2025laser, terhune2025photo}. Although its microscopic mechanism remains unresolved~\cite{schaden2025laser, derevianko2026colloquium}, the picture generally accepted at present couples the nucleus to bound electronic states, either through an electron bridge or through internal conversion into defect-derived in-gap states, with mobile charge carriers in the crystal participating in the process~\cite{guan2025x, elskens2026exploring}. Identifying those states requires the defect spectroscopy that we provide here.

The tetravalent dopant is a charged impurity, and it therefore requires charge compensation and introduces electronic states within the bandgap~\cite{takatori2025xafs}, whose calculated positions range from about 10~eV~\cite{Dessovic2014, nickerson2020nuclear, nickerson2021driven} down to 7.5~eV~\cite{morgan2025IC}. Efficient quenching, however, requires considerably more than the mere existence of such a state. The state must additionally supply occupied-to-empty spectral weight near the isomer energy $E_{\mathrm{iso}} = 8.3557$\,eV and must couple sufficiently strongly to the nucleus; a state that is strongly detuned from $E_{\mathrm{iso}}$, weakly coupled, or unfavorably occupied will contribute only weakly, since the internal-conversion rate is not determined by the level energy alone~\cite{elskens2026exploring, morgan2025IC, nalikowski2025fluorine}. The quantity of interest is therefore precisely the trap spectroscopy reported in this work.

Our measurements address this question in three ways. First, the glow curve locates eleven discrete trap levels with activation energies between $0.115(2)$ and $0.535(1)$\,eV, together with their associated emission bands (Tables~\ref{peaks} and~\ref{bands}). To our knowledge, this constitutes the first systematic determination of trap depths in $^{229}$Th:CaF$_2$, providing an experimentally determined trap-level spectrum rather than only qualitative evidence for the presence of defects. Second, the three afterglow components, with characteristic times near 1, 10, and 100~s [Fig.~\ref{afterglow}(g)], define the timescale on which trapped carriers are returned to the lattice and therefore remain available for capture at a thorium site, which is precisely what the carrier-mediated model requires~\cite{elskens2026exploring}. Third, the thermoluminescence and afterglow yields respond to the same temperature regions [Figs.~\ref{2024.10:CCD-TL}(c) and~\ref{afterglow}(f)], indicating that both draw on a common reservoir of trapped carriers; consequently, the afterglow, rather than the prompt scintillation, constitutes the optical fingerprint of the population that mediates quenching.

\subsection{Carrier trapping versus quenching: a quantitative link}
The trap spectrum determined above acquires a specific physical meaning when it is compared with the temperature dependence of x-ray-induced quenching measured on the same X2 crystal~\cite{guan2025x}. In that work the quenching rate of an individual isomer is expressed as $W_q = n^{\mathrm{e^-}}_{\mathrm{qch}}\,\sigma_{\mathrm{qch}}\,v(T)$, where $n^{\mathrm{e^-}}_{\mathrm{qch}}$ is the density of electrons capable of relaxing the isomer, $\sigma_{\mathrm{qch}}$ is the quenching cross section, and $v(T)$ is the mean diffusion velocity of the thermalized electrons. The central assumption of the model is that only mobile carriers can quench, so that

\begin{equation}
n^{\mathrm{e^-}}_{\mathrm{qch}}(T) = n_{\mathrm{e^-}} - n^{\mathrm{e^-}}_{\mathrm{trap}}(T),
\label{eq:nqch}
\end{equation}

in which $n_{\mathrm{e^-}}$ denotes the total density of migrating electrons set by the x-ray flux and $n^{\mathrm{e^-}}_{\mathrm{trap}}(T)$ the density of those captured by lattice traps. The competing process is therefore not the destruction of carriers but rather their \textit{storage}: once a carrier has been captured by a lattice trap, it is no longer available for capture at a thorium site. Within the same model the isomer yield is given by
$\mathrm{N_0}(T) = \varepsilon R\,/\{W_0 + [n_{\mathrm{e^-}} - n^{\mathrm{e^-}}_{\mathrm{trap}}(T)]\,\sigma_{\mathrm{qch}}\,v(T)\}$,
and because the afterglow intensity is proportional to $n^{\mathrm{e^-}}_{\mathrm{trap}}$, the afterglow and the isomer yield are predicted to share the same temperature dependence.

Two such features were indeed observed in Ref.~\cite{guan2025x}. First, the irradiation lifetime $\tau_{\mathrm{ir}}$ exhibits an unexpected maximum at \SI{-60}{\celsius}, corresponding to a quenching rate that is reduced relative to the neighboring temperatures, and the isomer yield $\mathrm{N_0}$ is enhanced at the same temperature. Second, $\mathrm{N_0}$ decreases from \SI{-60}{\celsius} to \SI{-80}{\celsius}; at \SI{-80}{\celsius} the afterglow yield is small, which by the model above implies a low $n^{\mathrm{e^-}}_{\mathrm{trap}}$ and hence a larger $n^{\mathrm{e^-}}_{\mathrm{qch}}$, so that the yield is suppressed. Below \SI{-80}{\celsius}, $\mathrm{N_0}$ rises again as the temperature is lowered further. Both features therefore reflect how many carriers are held in traps at a given temperature, and consequently how many remain available to quench the nucleus.

The thermoluminescence data presented here supply the missing half of this picture, because they locate directly the temperatures at which trapped carriers are released back into the lattice. Two glow peaks bracket the temperature range of interest: P5 at \SI{-55.6(2)}{\celsius}, with an activation energy of $0.372(3)$\,eV, and P4 at \SI{-105.6(1)}{\celsius}. The maximum of $\tau_{\mathrm{ir}}$ and $\mathrm{N_0}$ near \SI{-60}{\celsius} coincides with P5 within the experimental uncertainty, whereas the minimum near \SI{-80}{\celsius} lies between the P4 and P5 thermal-release regimes. Thus, the anomalous temperature dependence of the isomer yield occurs in the same temperature range spanning the P4- and P5-associated thermal-release regimes. Together with the observed temperature dependence of the afterglow, this correspondence supports the carrier-trapping picture of Ref.~\cite{guan2025x}. We emphasize, however, that the thermoluminescence intensity reflects the release and recombination of trapped carriers rather than the trap occupancy itself; a quantitative relation between the glow curve and the steady-state carrier density $n^{\mathrm{e^-}}_{\mathrm{trap}}(T)$ under x-ray irradiation would require a kinetic model of carrier capture and release.

Two consequences follow from this correspondence. First, the isomer yield is non-monotonic in temperature even though both the quenching cross section and the carrier mobility vary monotonically, because the yield is governed by the competition between carrier trapping and capture at the thorium site. The temperature range over which the isomer yield changes non-monotonically coincides with the distinct thermal-release regimes identified by the glow curve. Second, and more important for clock operation, the conditions that maximize the number of excitable nuclei are not necessarily those that minimize the optical background. The \SI{-60}{\celsius} region, where $\mathrm{N_0}$ is largest, lies near the onset of the thermal-release regime associated with P5--P7, with P5 peaking near \SI{-56}{\celsius}. Operation at higher temperatures suppresses the luminescence background while progressively releasing carriers from these traps. A quantitative treatment of this trade-off, however, requires not only the trap-level spectrum reported here but also the temperature-dependent balance between carrier trapping and release under irradiation.

A structural assignment of these traps to specific defect configurations cannot, however, be made from luminescence data alone, particularly because thorium is known to occupy several distinct sites in CaF$_2$~\cite{hiraki2026laser, morawetz2026continuous, rehn2026defect}. Since both the site population and the quenching efficiency are site-dependent~\cite{hiraki2026laser}, the trap spectrum reported here should be regarded as the superposition of the contributions of these sites; a site-resolved luminescence measurement on a crystal of known composition is therefore the natural next step. 


\subsection{Implications for solid-state nuclear clock operation}

The luminescence characterized in this work determines the optical background of a $^{229}$Th:CaF$_2$ clock. This background enters clock operation through the signal-to-noise ratio rather than as a frequency shift, and its relative weight depends strongly on the readout scheme employed. The thorium-229 clocks demonstrated so far use absorption readout, for which the signal photon rate exceeds that of fluorescence detection by three orders of magnitude~\cite{de2026thorium,huang2026nuclear}; moreover, the first closed-loop device operates at room temperature (294.7~K) with 10~Hz on/off differential detection and reports no effect of $\sim$0.1~K temperature fluctuations on the clock~\cite{de2026thorium}, so that in this configuration the luminescence background is not a limiting factor. Fluorescence detection, which is the natural choice for VUV counting at low excitation rates~\cite{ooi2025frequency,higgins2025temperature}, is instead exposed directly to the background channels quantified here.

Two consistency checks connect our measurements with reported clock operation. First, the STE emission has been used as an in-situ monitor of the excitation laser power, because it scales with laser intensity rather than with laser frequency~\cite{ooi2025frequency}; our finding that this channel is quenched by more than two orders of magnitude between room temperature and \SI{100}{\celsius} [Fig.~\ref{veto-PMT-RL}(f)] provides the calibration needed to use it as a power reference over a wide temperature range. Second, an increase of the background count rate persisting after a frequency scan, which distorts the resonance lineshape and is corrected by subtracting a linear baseline, has been reported in the same experiment~\cite{ooi2025frequency}; this is the signature of the afterglow characterized here, whose decay we resolve into components near 1, 10, and 100~s [Fig.~\ref{afterglow}(g)].

The glow peaks (Table~\ref{peaks}), their associated emission bands (Table~\ref{bands}), and the afterglow components reported here have, to our knowledge, no systematic counterpart in the literature for $^{229}$Th:CaF$_2$. They therefore provide the quantitative basis for background-suppression strategies that have so far been discussed only in general terms, such as the use of veto detection or of crystals with a lower intrinsic background~\cite{ooi2025frequency}. With this in mind, our results suggest the following operational considerations.

\begin{enumerate}
    \item \textbf{Operating temperature.} Two regimes are available, governed by different criteria. To minimize the optical background, operation near \SI{100}{\celsius} suppresses the STE radioluminescence by more than two orders of magnitude [Fig.~\ref{veto-PMT-RL}(f)] and shortens the afterglow from minutes to seconds [Fig.~\ref{afterglow}(g)]. To minimize the nuclear transition frequency shift, operation at the zero-shift temperature $T_0 = 196(5)$\,K (\SI{-77}{\celsius}) is preferred~\cite{ooi2025frequency}, provided that the crystal is enclosed in a cryogenic copper shield to prevent ice formation [Fig.~\ref{veto-PMT-RL}(g)]. These two criteria are independent of each other, since $T_0$ follows from the temperature dependence of the quadrupole-split nuclear transition and is unrelated to the luminescence background. In both regimes the Cherenkov background is stable and can be referenced out.

    \item \textbf{Pre-conditioning (routine).} A 10~min hold at \SI{100}{\celsius} between measurement cycles releases the carriers trapped in the shallow defects corresponding to peaks P8--P11 (Table~\ref{peaks}), thereby resetting the thermoluminescence background. This is particularly important after operation below \SI{-80}{\celsius}, where carriers captured in the deeper traps associated with P5--P7 can remain trapped.

    \item \textbf{Full annealing (periodic).} Prolonged self-irradiation accumulates permanent defects, which contribute the broadband emission of bands B4--B6. A full annealing step at \SI{400}{\celsius} for several hours restores the crystal to a nearly defect-free state. On the basis of the accumulation rate set by the thorium activity, we suggest a full anneal every $\sim$1--2 weeks of continuous operation.

    \item \textbf{Vacuum and ice management.} The chamber pressure should be maintained at or below $10^{-4}$~Pa. Below \SI{-100}{\celsius}, the copper shield with its MgF$_2$ window is essential, since residual gas would otherwise condense on the crystal and absorb both the 148~nm isomer signal and the Cherenkov reference photons.
\end{enumerate}

Taken together, our characterization of the luminescence of $^{229}$Th:CaF$_2$ both identifies the defect dynamics that underlie isomer quenching and quantifies the background channels that limit the signal-to-noise ratio of fluorescence-based clocks. The operating procedures outlined above---a moderate operating temperature, routine thermal cycling, and periodic annealing---provide a starting point for background control that is compatible with frequency-optimized operation at the zero-shift temperature.


\section{Conclusions}

Photoluminescence and radioluminescence in $^{229}$Th:CaF$_2$ share a single mechanism, namely the recombination of nearest-neighbor F and H centers, i.e., STE scintillation; the two differ only in the excitation, which is external for the former and arises from radioactive decay for the latter. In both cases the dominant emission is the STE light burst between 200 and 450~nm, peaking near 280~nm. This component is thermally quenched, with fitted activation energies of $0.38(1)$\,eV for the triplet and $0.10(1)$\,eV for the singlet component. The Cherenkov component (120--200~nm), by contrast, is largely insensitive to temperature.

Thermoluminescence reveals a series of glow peaks together with the corresponding emission bands. The peaks yield the activation energies required to release the trapped carriers, while the bands reflect the energy released upon recombination. Below room temperature the emission is concentrated in the short-wavelength bands that resemble STE scintillation; as the temperature rises, these bands are suppressed and longer-wavelength bands become dominant.

The afterglow depends strongly on temperature, and its yield tracks that of the thermoluminescence. Lifetimes are long where the thermoluminescence yield is low, as between \SI{-100}{\celsius} and \SI{-60}{\celsius}, whereas within the regions of strong thermoluminescence the lifetime increases as the temperature decreases. Finally, the trap depths determined here correlate with the temperature dependence of x-ray-induced isomer quenching, supporting a common defect-related origin of the optical and nuclear observables of this material.

\begin{acknowledgements}
    {This work was supported by JSPS KAKENHI Grant Numbers JP21H04473, JP23K13125, JP24K00646, JP24H00228, JP24KJ0168. This work was also supported by JST CREST Grant No. JPMJCR24I6 and by JSPS Bilateral Joint Research Projects No. 120222003. This work has been funded by the European Research Council (ERC) under the European Union’s Horizon 2020 research and innovation programme (Grant Agreement No. 856415,  No. 101266822 and No.  101087184) and the Austrian Science Fund (FWF) [Grant DOI: 10.55776/F1004, 10.55776/KIN4315625]. This work is also supported by the Strategic Priority Research Program of the Chinese Academy of Science (Grant No. XDB0920000) and the Natural Science Foundation of China (Grant No. 12604402).}
\end{acknowledgements}


\bibliographystyle{apsrev4-2}

\bibliography{lumi}

\end{document}